\documentclass[12pt]{article}
\usepackage{amsmath}
\usepackage{color}
\def\be{\begin{equation}}
\def\ee{\end{equation}}
\def\arr{\begin{array}{rll}}
\def\ea{\end{array}}
\def\bea{\begin{eqnarray}}
\def\eea{\end{eqnarray}}

\def\N2{$N{=}2$}

\def\>{\rangle}
\def\<{\langle}
\def\+{\dagger}
\def\={\ =\ }

\begin{document}
\renewcommand{\thefootnote}{\fnsymbol{footnote}}
\begin{titlepage}
\setcounter{page}{0}
\begin{flushright}
LMP-TPU--1/14  \\
\end{flushright}
\vskip 1cm
\begin{center}
{\LARGE\bf Conformal Newton--Hooke symmetry }\\
\vskip 0.5cm
{\LARGE\bf of Pais--Uhlenbeck oscillator}\\
\vskip 1cm
$
\textrm{\Large K. Andrzejewski\ }^{a}, ~ \textrm{\Large A. Galajinsky\ }^{b,c}, ~ \textrm{\Large J. Gonera\ }^{a} ~ \textrm{\Large and \ } ~ \textrm{\Large I. Masterov \ }^{b}
$
\vskip 0.7cm
${}^{a}$
{\it Department of Computer Science, Faculty of Physics and Applied Informatics \\
University of \L \'od\'z, Pomorska 149/153, 90-236 \L \'od\'z, Poland}
\vskip 0.5cm
${}^{b}$
{\it
Laboratory of Mathematical Physics, Tomsk Polytechnic University, \\
634050 Tomsk, Lenin Ave. 30, Russian Federation} \\
\vskip 0.5cm
${}^{c}$
{\it
Department of Physics, Tomsk State University, 634050 Tomsk, \\
Lenin Ave. 36, Russian Federation} \\
\vskip 0.4cm
{E-mails: k-andrzejewski@uni.lodz.pl, galajin@tpu.ru, \\
~ ~ jgonera@uni.lodz.pl, masterov@tpu.ru}
\end{center}
\vskip 1cm
\begin{abstract} \noindent
It is demonstrated that the Pais--Uhlenbeck oscillator in arbitrary dimension enjoys the $l$--conformal Newton--Hooke symmetry provided frequencies of oscillation form
the arithmetic sequence $\omega_k=(2k-1) \omega_1$, where $k=1,\dots,n$, and $l$ is the half--integer $\frac{2n-1}{2}$. The model is shown to be maximally superintegrable.
A link to $n$ decoupled isotropic oscillators is discussed and an interplay between the $l$--conformal Newton--Hooke symmetry and symmetries characterizing
each individual isotropic oscillator is analyzed.

\end{abstract}


\vskip 0.5cm


\end{titlepage}

\renewcommand{\thefootnote}{\arabic{footnote}}
\setcounter{footnote}0

\noindent
{\bf 1. Introduction}\\

\noindent

Nondegenerate higher derivative theories generically show up instability in classical dynamics and bring about violation of unitarity and/or trouble with ghosts in quantum theory \cite{W}.
The Pais--Uhlenbeck (PU) oscillator in one dimension \cite{PU} is arguably the most popular higher derivative mechanical system which attracted considerable interest in the past (for reviews and further references see \cite{S,AKM,PK}).

Recall that a Lagrangian system is called of degree $N$ if the Lagrangian density involves derivatives of dynamical variables up to the order $N$. Ostrogradski's method then implies that the corresponding
Hamiltonian is linear in $N-1$ canonical momenta. An immediate corollary is that at least $N-1$ of $2N$ functionally independent solutions of the classical equations of motion carry negative energy.
The PU oscillator is by no means an exclusion. Its energy is not bounded from below as half of its solutions carry negative energy. It is important to stress, however, that because the positive and negative energy modes do not interact with each other the classical model is stable (see Eq. (\ref{gs}) below). Only in the presence of interaction can the energy flow from one mode to another and the runaway solutions appear which signal the classical instability. An example is the external friction force studied in \cite{VN}. Note that in quantum theory the problem becomes more pronounced because the absence of the ground state within the framework of the conventional quantization scheme makes a physical interpretation of the quantum PU oscillator troublesome. Within the alternative approach (see, e.g., the discussion in \cite{AKM} and references therein) one can define a ground state but the appearance of ghosts is unavoidable. Thus, while the PU oscillator describes a stable dynamical system within the context of classical mechanics, it does not yield a physically viable quantum mechanical model.

Despite being rather popular, the classical PU oscillator and its multidimensional generalization do not seem to have been thoroughly investigated with regard to their symmetries. Yet, as was demonstrated in the original work \cite{PU},
the PU oscillator is dynamically equivalent to a set of decoupled harmonic oscillators. As is known since Niederer's work \cite{N}, the harmonic oscillator is invariant under the conformal group $SO(2,1)$. Its multidimensional generalization exhibits the $l=\frac 12$ conformal Newton--Hooke (NH) symmetry \cite{G}. Then it is natural to wonder whether the (multidimensional) PU oscillator is conformal invariant and which is its full symmetry group.

In Ref. \cite{GK} its was conjectured that the symmetry is described by the $l$--conformal NH group \cite{NOR,GM}. However, the explicit example of $l=\frac 32$ studied in \cite{GM1} shows that only for the special case that  frequencies are related via $\omega_2=3\omega_1$ is the PU oscillator conformal invariant. The principal objective of this work is to demonstrate that the classical PU oscillator in arbitrary dimension enjoys the $l$--conformal NH symmetry provided frequencies of oscillation form the arithmetic sequence $\omega_k=(2k-1) \omega_1$, where $k=1,\dots,n$, and $l$ is the half--integer $\frac{2n-1}{2}$.

The investigation of the maximal kinematical invariance group of the multidimensional PU oscillator links nicely with the recent extensive study of nonrelativistic conformal algebras and their dynamical realizations
\cite{GM}--\cite{AKS}. Although the $l$--conformal extension of the NH algebra is known for a long time \cite{NOR} \footnote{The flat space limit of the $l$--conformal NH algebra in \cite{NOR} does not yield the $l$--conformal Galilei algebra. This shortcoming was overcame in \cite{GM}.}, its dynamical realizations remain almost completely unexplored.
In a recent work \cite{GM1}, the method of nonlinear realizations was applied to the $l$--conformal NH algebra to construct a dynamical system
without higher derivative terms in the equations of motion. The present paper aims to provide another dynamical realization in the context of a consistent higher derivative theory.

The work is organized as follows. In the next section we briefly remind the basic facts about the $l$--conformal NH algebra. Sect. 3 is devoted to a systematic derivation of the maximal kinematical invariance group of the classical PU oscillator in arbitrary dimension. In particular, it is demonstrated that the model accommodates the $l$--conformal NH symmetry provided frequencies of oscillation form the arithmetic sequence $\omega_k=(2k-1) \omega_1$, where $k=1,\dots,n$, and $l$ is the half--integer $\frac{2n-1}{2}$. In Sect. 4 it is shown that the same result is attained if one applies Niederer's transformation \cite{N} to a free
particle obeying a higher derivative equation of motion which is invariant under the $l$--conformal Galilei group.
In Sect. 5 we discuss a description of the PU model in terms of decoupled isotropic oscillators and demonstrate that the model is maximally superintegrabile provided ratios of frequencies are rational numbers. The
case of the quartic PU oscillator which is linked to $l=\frac 32$ is discussed in detail. It is shown that, while the variables parametrizing distinct isotropic oscillators are decoupled in the Hamiltonian, they are nontrivially intertwined within other conserved charges which all together form a central extension of the $l$--conformal NH algebra under the Poisson bracket.
In the concluding Sect. 6 we summarize our results and discuss possible further developments.

 \vspace{0.5cm}

\noindent
{\bf 2. The $l$--conformal NH algebra}\\

\noindent
The $l$--conformal NH algebra includes the generators of time translations, dilatations, special conformal transformations, spatial rotations,
spatial translations, Galilei boosts
and accelerations. Denoting them by $(H,D,K, M_{ij}, C^{(p)}_i)$, where $i=1,\dots,d$ is a spatial index and $p=0,1,\dots, 2l$ with a half--integer $l$,
one has the structure relations \cite{NOR,GM}
\begin{align}\label{algebra}
&
[H,D]=H \mp \frac{2}{R^2} K, && [H,C^{(p)}_i]=p C^{(p-1)}_i \pm \frac{(p-2l)}{R^2} C^{(p+1)}_i,
\nonumber\\[2pt]
&
[H,K]=2 D, && [D,K]=K,
\nonumber\\[2pt]
&
[D,C^{(p)}_i]=(p-l) C^{(p)}_i, && [K,C^{(p)}_i]=(p-2l) C^{(p+1)}_i,
\nonumber\\[2pt]
&
[M_{ij},C^{(p)}_k]=-\delta_{ik} C^{(p)}_j+\delta_{jk} C^{(p)}_i, && [M_{ij},M_{kl}]=-\delta_{ik} M_{jl}-\delta_{jl} M_{ik}+
\delta_{il} M_{jk}+\delta_{jk} M_{il}.
\end{align}
The instances of $p=0$ and $p=1$ in $C^{(p)}_i$ correspond to space translations and the Galilei boosts, while higher values of $p$ are related to accelerations.
The constant $R$ is called the characteristic time and $\pm\frac{1}{R^2}$ is interpreted as a nonrelativistic cosmological constant (for a more detailed discussion see \cite{BAC,GP}). The upper/lower sign in the first line in (\ref{algebra}) corresponds to a negative/positive cosmological constant. A realization of the algebra in nonrelativistic spacetime can be found in \cite{GM}.
In what follows we consider only a negative cosmological constant as it generically leads to the stable classical dynamics.

In arbitrary dimension and for a half--integer $l$ the algebra admits a central extension \cite{GM}
\be
[C^{(p)}_i,C^{(m)}_j]={(-1)}^p p! m!  \delta_{ij} \delta_{p+m,2l} \alpha,
\ee
where $\alpha$ is an arbitrary constant. In dynamical realizations the latter is linked to physical parameters of a system such as the mass.

The limit of a vanishing cosmological constant yields the $l$--conformal Galilei algebra \cite{NOR}. The latter can be also obtained by
a formal linear change of the basis $H \quad \rightarrow \quad H\mp \frac{1}{R^2} K$,
where the upper/lower sign corresponds to a negative/positive cosmological constant. As far as dynamical realizations are concerned, the use of the new basis implies the change of the Hamiltonian
which alters the dynamics. By this reason, the $l$--conformal NH algebra and its Galilei counterpart are usually considered separately. It should also be remembered that $R$ is a dimensionful constant which may not be at one's disposal in a flat nonrelativistic spacetime.

\vspace{0.5cm}

\noindent
{\bf 3. Symmetries of the multidimensional PU oscillator}\\

\noindent
Let us consider the equation of motion of the PU oscillator in arbitrary dimension
\bea\label{PU}
&&
\prod_{k=1}^{n}\left(\frac{d^2}{dt^2}+\omega_k^2\right)x_i(t)=\sum_{k=0}^{n}\sigma_k^{n}\frac{d^{2k}}{dt^{2k}}x_i(t)=0,
\\[2pt]
&&
\sigma_n^{n}=1,  \qquad \sigma_k^{n}=\sum_{i_1 <i_2<\dots < i_{n-k}}^n \omega_{i_1}^2 \dots \omega_{i_{n-k}}^2,
\nonumber
\eea
where $i=1,\dots,d$ and, for definiteness, we assume that $0<\omega_1 < \omega_2< \dots < \omega_n$. The form of the differential operator which enters the left hand side of (\ref{PU}) prompts one to find the general solution
\be\label{gs}
x_i (t)=\sum_{k=1}^n (\alpha_i^k \cos{(\omega_k t)}+\beta_i^k \sin{(\omega_k t)}),
\ee
where $\alpha_i^k$ and $\beta_i^k$ are constants of integration. Our objective in this section is to find out under which circumstances (\ref{PU}) holds invariant under the standard infinitesimal transformations
adopted in classical mechanics
\bea\label{inftr}
t'=t+\epsilon\psi(t),\qquad x_i'(t')=x_i(t)+\epsilon\phi_i(t,x),
\eea
where $\epsilon$ is an infinitesimal parameter and $\psi(t)$, $\phi_i(t,x)$ are the functions to be determined.

The invariance of the equation\footnote{Throughout this work, unless explicitly stated otherwise, summation over repeated indices is understood.}
\bea\label{inv}
&&
\prod_{k=1}^{n}\left(\frac{d^2}{dt'^2}+\omega_k^2\right)x_i'(t')=(\delta_{ij}+\epsilon \lambda_{ij} (t,x)) \prod_{k=1}^{n}\left(\frac{d^2}{dt^2}+\omega_k^2\right)x_j(t)
\eea
where $\lambda_{ij} (t,x)$ is an invertible matrix function to be fixed below, yields\footnote{The simplest way to derive (\ref{peq}) is to represent the operator
$\frac{d}{d t'}=(1-\epsilon \dot\psi) \frac{d}{d t}$ in the form $\frac{d}{d t'}=e^{A} \frac{d}{d t} e^{-A}$, $A=\epsilon \psi(t) \frac{d}{d t}$, which yields $\prod_{k=1}^{n}\left(\frac{d^2}{dt'^2}+\omega_k^2\right)=
\prod_{k=1}^{n}\left(\frac{d^2}{dt^2}+\omega_k^2\right)+[A, \prod_{k=1}^{n}\left(\frac{d^2}{dt^2}+\omega_k^2\right)]$ and then to make use of the identity $[\frac{d^n}{dt^n},f(t)]=\sum_{k=0}^{n-1} C_n^k f^{(n-k)} (t) \frac{d^{k}}{d t ^{k}}$.}
\bea\label{peq}
\sum_{k=0}^{n}\sigma_k^{n}\phi_i^{(2k)}(t,x)-\sum_{k=1}^{n}\sigma_k^{n}  \sum_{p=1}^{2k}C_{2k}^{p-1}\psi^{(2k-p+1)} x_i^{(p)}(t)=\lambda_{ij} (t,x) \sum_{k=0}^{n}\sigma_k^{n}x_j^{(2k)}(t).
\eea
As usual, a superscript in braces designates the number of derivatives with respect to the temporal coordinate, e.g. $x_i^{(p)}(t)=\frac{d^p x_i (t) }{dt^p}$, ${\dot x}_i(t)=\frac{d x_i (t) }{dt}$.

The form of the constraint (\ref{peq}) implies
that $\phi_i(t,x)$ is a linear function of the variable $x$
\bea\label{phi}
\phi_i(t,x)=a_{ij}(t) x_j(t)+b_i(t).
\eea
Substituting (\ref{phi}) into (\ref{peq}), at the zeroth and first orders in $x$ one finds
\bea\label{lb}
&&
\sum_{k=0}^{n}\sigma_k^{n} b_i^{(2k)}=0, \qquad \lambda_{ij}=\frac{1}{\sigma_{0}^{n}} \sum_{k=0}^{n}\sigma_k^{n}a_{ij}^{(2k)},
\eea
while the rest involves the derivatives of $x$
\bea\label{rest}
&&
\sum_{k=1}^{n}\sigma_k^{n}\sum_{p=1}^{2k}\left(C_{2k}^{p}a_{ij}^{(2k-p)}-C_{2k}^{p-1}\psi^{(2k-p+1)}\delta_{ij}-\lambda_{ij}\delta_{p,2k}\right)x_j^{(p)}=0.
\eea

The rightmost equation in (\ref{lb}) relates $\lambda_{ij}$ to $a_{ij}$, while  $b_i(t)$ obeys the same equation as $x_i(t)$ and reads
\be\label{parb}
b_i (t)=\sum_{k=1}^n (\mu_i^k \cos{(\omega_k t)}+\nu_i^k \sin{(\omega_k t)}),
\ee
where $\mu_i^k$ and $\nu_i^k$, are arbitrary constants which, being multiplied by $\epsilon$ in (\ref{inftr}), yield infinitesimal parameters of the corresponding transformations. The transformation
$x'_i(t)=x_i(t)+\epsilon b_i(t)$ thus maps a solution (\ref{gs}) of (\ref{PU}) into another one which corresponds to changing initial conditions for the Cauchy problem.

Let us turn to Eq. (\ref{rest}). Gathering the terms which involve even and odd number of derivatives acting on $x$, one gets\footnote{The conventional properties of the double sum $\sum_{k=1}^{n}\sum_{p=1}^{2k}g(k,p)=\sum_{k=1}^{n}\sum_{p=1}^{k} (g(k,2p)+g(k,2p-1))$ and $\sum_{k=1}^{n}\sum_{p=1}^{k}g(k,p)=\sum_{p=1}^{n}\sum_{k=p}^{n}g(k,p)$, where $g(k,p)$ is an arbitrary function, prove to be helpful.}
\bea
&&
\sum_{p=1}^{n}\sum_{k=p}^{n}\sigma_k^{n} \left( \left[C_{2k}^{2p}a_{ij}^{(2k-2p)}-C_{2k}^{2p-1}\psi^{(2k-2p+1)}\delta_{ij}-\lambda_{ij}\delta_{p,k}\right]x_j^{(2p)}+
\right.
\nonumber\\[2pt]
&&
\qquad \qquad
\left.
+\left[C_{2k}^{2p-1}a_{ij}^{(2k-2p+1)}-C_{2k}^{2p-2}\psi^{(2k-2p+2)}\delta_{ij}\right]x_j^{(2p-1)}\right)=0,
\eea
which implies
\bea\label{constr}
&&
\sum_{k=p}^{n}\sigma_k^{n}\left(C_{2k}^{2p-1}a_{ij}^{(2k-2p+1)}-C_{2k}^{2p-2}\psi^{(2k-2p+2)}\delta_{ij}\right)=0,
\nonumber\\[2pt]
&&
\sum_{k=p}^{n}\sigma_k^{n}\left(C_{2k}^{2p}a_{ij}^{(2k-2p)}-C_{2k}^{2p-1}\psi^{(2k-2p+1)}\delta_{ij}\right)=\lambda_{ij}\sigma_p^{n}, \qquad p=1,\dots,n.
\eea

Choosing $p=n$ in the first line in (\ref{constr}), one derives the differential equation
\be
{\dot a}_{ij}(t)-\frac{2n-1}{2} \psi^{(2)} (t) \delta_{ij}=0,
\ee
which is readily integrated to yield
\be\label{aij}
a_{ij}(t)=\frac{2n-1}{2} {\dot\psi} (t) \delta_{ij}+a^0_{ij},
\ee
where $a^0_{ij}$ is a constant matrix. Being multiplied by $\epsilon$ in (\ref{inftr}), the latter generates the infinitesimal $GL(d,R)$ transformation, which is obviously a symmetry of (\ref{PU}).

Setting $p=n$ and $p=n-1$ in the second line in (\ref{constr}) and taking into account (\ref{aij}), one then gets
\bea\label{om}
&&
\psi^{(3)}(t)+\tilde\omega^2 {\dot\psi}(t)=0, \qquad \tilde\omega^2=\frac{12}{n(4n^2-1)} \sum_{k=1}^n \omega_k^2,
\\[2pt]
&&
\lambda_{ij}=-\frac{2n+1}{2} {\dot\psi} (t) \delta_{ij}+a^0_{ij}.
\nonumber
\eea
Thus the three equations mentioned above fix $\psi(t)$
\be\label{abc}
\psi(t)=a+ b \sin{(\tilde\omega t)}+c\cos{(\tilde\omega t)},
\ee
where $a$, $b$, and $c$ are arbitrary constants, and thereby determine also $a_{ij}(t)$ and $\lambda_{ij}(t)$. It is straightforward to verify that the generators\footnote{As usual, the generators of a global symmetry transformation are derived from $\psi(t)$ and $\phi_i(t,x)$ in
Eq. (\ref{inftr}). In order to derive $K$ in (\ref{HDK}), one chooses $\psi(t)$ in the form $\psi(t)=\cos{(\tilde\omega t)}-1$.} corresponding to the transformations (\ref{abc}) (cf. Eq. (3) in Ref. \cite{GM})
\bea\label{HDK}
&&
H=\partial_t, \qquad D=\frac{1}{\tilde\omega} \sin{(\tilde\omega t)} \partial_t+\frac{2n-1}{2} \cos{(\tilde\omega t)} x_i\partial_i,
\nonumber\\[2pt]
&&
K=-\frac{2}{\tilde\omega^2} (\cos{(\tilde\omega t)}-1) \partial_t+\frac{2n-1}{\tilde\omega} \sin{(\tilde\omega t)} x_i \partial_i,
\eea
where $\partial_t=\frac{\partial}{\partial t}$, $\partial_i=\frac{\partial}{\partial x_i}$,
obey the structure relations of the conformal algebra in one dimension $so(2,1)$, $a$, $b$, and $c$  being the parameters of the time translation, dilatation and the special conformal transformation, respectively.

The remaining equations in (\ref{constr}) along with the rightmost condition in (\ref{lb}) yield constraints on $\sigma^n_p$, i.e. on frequencies of oscillation $\omega_1,\dots,\omega_n$, which can be solved recursively to fix the admissible values.
Because the restrictions turn out to be highly nonlinear, we first give a simpler symmetry argument that frequencies form the arithmetic sequence
\be\label{apr}
\omega_k=(2k-1) \omega_1,
\ee
where $k=1,\dots,n$,
and then verify that the rest in (\ref{constr}) and the rightmost equation in (\ref{lb}) are identically satisfied.

Given a solution $x_i(t)$ of the equation of motion, let us require the transformed function
\be
x'_i(t)=x_i(t)+\frac{2n-1}{2} \epsilon \dot\psi(t) x_i(t)-\epsilon \psi(t) {\dot x}_i(t)
\ee
to be a new solution of (\ref{PU}). For definiteness, let us focus on the dilatation transformation generated by $\psi(t)=\sin{(\tilde\omega t)}$ with $\tilde\omega$ given in (\ref{om}). Taking into account (\ref{gs}) and the standard properties of trigonometric functions, one readily gets
\bea
&&
x'_i(t)=\sum_{k=1}^n (\alpha_i^k \cos{(\omega_k t)}+\beta_i^k \sin{(\omega_k t)})+
\nonumber\\[2pt]
&&
\quad
+\frac{\epsilon}{2}\sum_{k=1}^n  \alpha_i^k  \left(\left[\frac{(2n-1)}{2} \tilde\omega-\omega_k\right] \cos{(\omega_k+\tilde\omega)t} +\left[\frac{(2n-1)}{2} \tilde\omega+\omega_k\right] \cos{(\omega_k-\tilde\omega)t}\right)+
\nonumber\\[2pt]
&&
\quad
+\frac{\epsilon}{2} \sum_{k=1}^n  \beta_i^k  \left(\left[\frac{(2n-1)}{2} \tilde\omega-\omega_k\right] \sin{(\omega_k+\tilde\omega)t} +\left[\frac{(2n-1)}{2} \tilde\omega+\omega_k\right] \sin{(\omega_k-\tilde\omega)t}\right).
\eea
Since both $\omega_k$ and $\tilde\omega$ are positive, the only way to generate a new solution is to require
\be\label{cono}
\omega_k+\tilde\omega=\omega_{k+1}, \qquad \frac{(2n-1)}{2} \tilde\omega-\omega_n=0,
\ee
where $k=1,\dots,n-1$, which immediately yields
\be\label{cono1}
\tilde\omega=2\omega_1.
\ee
Note that the latter relation and Eq. (\ref{cono}) is consistent with the definition of $\tilde\omega$ in (\ref{om}). At this stage, it is straightforward to verify that the remaining equations in
(\ref{constr}) and the rightmost equation in (\ref{lb}) are identically satisfied.

Note that an alternative possibility to derive (\ref{apr}) is to compute the algebra of the conformal transformations revealed above and the transformations with the vector parameters (\ref{parb}). It turns out that for generic values of frequencies the algebra does not close. Requiring the closure of the algebra, one precisely reproduces the restrictions (\ref{cono}).

Above we have considered the dilatation transformation. The special conformal transformation can be treated likewise and leads to the same result (\ref{cono}).

Having fixed frequencies, one can rewrite the generators of the transformations with the parameters (\ref{parb}) in the equivalent form
\be
\label{ces}
C^{(p)}_i={ \left( \frac{2}{\tilde\omega} \tan{\frac{\tilde\omega t}{2}}\right)}^p {\left( \cos{\frac{\tilde\omega t}{2}} \right)}^{2l} \partial_i,
\ee
where $p=0,\dots,2l$,
and verify that these along with $H$, $D$ and $K$ in (\ref{HDK}) they do obey the structure relations of the $l$--conformal NH algebra \cite{GM} ~\footnote{As we have seen above, the transformations with the parameters $a^0_{ij}$ in (\ref{aij}) generate the general linear group $GL(d,R)$. In particular, the antisymmetric part $a_{[ij]}$ is responsible for spatial rotations. For the equation of motion (\ref{PU}) the $l$--conformal NH algebra is thus extended by the transformations generated by the symmetric part $a_{(ij)}$. In general, such transformations are discarded as the Lagrangian formulation for (\ref{PU}) enjoys only the rotation symmetry.} with $l=\frac{2n-1}{2}$.

We thus conclude that
the multidimensional Pais--Uhlenbeck oscillator enjoys the $l$--conformal NH symmetry for the special case that
frequencies of oscillation form the arithmetic sequence $\omega_k=(2k-1) \omega_1$ with $k=1,\dots,n$ and $l$ is the half--integer $\frac{2n-1}{2}$.

\vspace{0.5cm}

\noindent
{\bf 4. Niederer's transformation}\\

\noindent

In the previous section we revealed the NH symmetry in the PU oscillator by a direct computation. Let us demonstrate that the same result is attained if one applies an analogue of
Niederer's transformation \cite{N} to a free particle obeying the higher derivative equation of motion
\be\label{fe}
\frac{d^{2n} x_i (t)}{d t^{2n}}=0.
\ee

As is known \cite{GK,AG}, Eq. (\ref{fe}) holds invariant under the action of the $l$--conformal Galilei group\footnote{To be more precise, (\ref{fe}) is invariant under the $l$--conformal Galilei group extended by the extra transformations of the form $\delta x_i=a_{(ij)} x_j$, where $a_{(ij)}$ is a symmetric matrix, which extend rotations to the full general linear group.} with the half--integer $l=\frac{2n-1}{2}$.
A conventional realization of the corresponding generators reads
\bea\label{gen}
&&
H=\partial_t, \qquad \qquad D=t \partial_t+l x_i \partial_i, \qquad \qquad K=t^2 \partial_t+2 l t x_i \partial_i,
\nonumber\\[2pt]
&&
C^{(p)}_i=t^p \partial_i, \qquad M_{ij}=x_i \partial_j-x_j \partial_i,
\eea
where $p=0,1,\dots, 2l$ and  $i=1,\dots,d$.

As was mentioned above, an analogue of the $l$--conformal Galilei algebra in the presence of a universal cosmological repulsion or attraction is the $l$--conformal
NH algebra. For the case of a negative cosmological constant the generators have the form  \cite{GM}
\bea\label{gener}
&&
H=\partial_t, \qquad
D=\frac 12 R \sin{(2t/R)} \partial_t+l \cos{(2t/R)} x_i \partial_i,
\nonumber\\[2pt]
&&
K=-\frac 12 R^2 (\cos{(2t/R)}-1) \partial_t+l R \sin{(2t/R)} x_i \partial_i,
\nonumber\\[2pt]
&&
C^{(p)}_i=R^p {(\tan{(t/R)})}^p {(\cos{(t/R)})}^{2l} \partial_i, \qquad  M_{ij}=x_i \partial_j-x_j \partial_i.
\eea

Niederer's transformation relates the motion of a free particle to a half--period of the harmonic oscillator \cite{N}. Its analogue which
links (\ref{gen}) to  (\ref{gener})~\footnote{
Recall that the $l$--conformal Galilei algebra and its NH counterpart are isomorphic.
The linear change of the basis $H \quad \rightarrow \quad H - \frac{1}{R^2} K$ in the $l$--conformal NH algebra yields the $l$--conformal Galilei algebra \cite{GM}.
When applying Niederer's transformation, the redefinition of the time translation generator should be taken into account.} reads \cite{GM}
\be\label{nied}
t'=R \tan{(t/R)}, \qquad x'_i={(\cos{(t/R)})}^{-2l} x_i.
\ee
Here the prime denotes the coordinates parameterizing a flat space.

Let us
apply the transformation (\ref{nied}) to the equation
(\ref{fe}) in a flat spacetime which is invariant under the $l$--conformal Galilei group. By construction, the resulting equation will enjoy the $l$--conformal NH symmetry.
A straightforward computation yields
\be
\prod_{k=1}^{l+\frac 12} \left(\frac{d^2}{d t ^2}+\frac{{(2k-1)}^2}{R^2} \right) x_i(t)=0,
\ee
which is a variant of the multidimensional PU oscillator ($\omega_1=\frac{1}{R}$) considered above.
Thus, Niederer's transformation fixes frequencies unambiguously and reproduces the result in the preceding section.

\vspace{0.5cm}

\noindent
{\bf 5. Decoupled oscillators, superintegrability and NH symmetry}\\

\noindent

The form of the general solution (\ref{gs}) suggests that the PU oscillator is dynamically equivalent to a set of decoupled isotropic oscillators with frequencies $\omega_1, \dots, \omega_n$.
As was demonstrated in the original work \cite{PU}, there exist canonical variables
in which the Hamiltonian of the PU oscillator in $d=1$ turns into the direct sum of harmonic oscillators with alternating sign.
The argument in \cite{PU} is readily generalized to arbitrary dimension. Introducing the new variables\footnote{In this section we switch to the vector notation and omit spatial indices. As follows from the condition $\omega_1<\dots<\omega_n$,
the constants $\rho _k$ are positive. In deriving (\ref{lagr}), the identity
$\rho_1 \Pi_2 \Pi_3 \dots \Pi_n-\Pi_1 \rho_2 \Pi_3 \dots \Pi_n+\dots+{(-1)}^{n-1} \Pi_1 \Pi_2 \Pi_3 \dots \rho_n=1$, where $\Pi_k=\frac{d^2}{dt^2}+\omega_k^2$, proves to be helpful.}
\be
\label{coord}
\vec{x}_i = \sqrt{\rho _i}\prod_{k \neq i} \left(\frac{d^2}{dt^2} + \omega _k^2\right)\vec x, \qquad \rho_i = \frac{{(-1)}^{i+1}}{\prod_{k \neq i}(\omega ^2_k -\omega ^2_i)},
\ee
where $i=1,\dots,n$,  one can bring the PU Lagrangian to that describing $n$ decoupled isotropic oscillators
\be
\label{lagr}
L= -\frac 12 \vec{x} \prod_{k=1}^{n}\left(\frac{d^2}{dt^2} +\omega^2_k\right) \vec{x}=\frac{1}{2} \sum_{k=1}^{n} {(-1)}^{k+1} \left( \dot{\vec{x}}_{k} ^2 - \omega _k^2\vec{x}_k^2\right),
\ee
where we discarded total derivative terms on the right hand side. The form of the corresponding Hamiltonian
\be
\label{free}
H=\frac 12 \sum_{k=1}^n {(-1)}^{k+1} (\vec p_k^2+\omega_k^2 \vec x_k^2),
\ee
drastically facilitates the analysis of superintegrability.

For a single isotropic oscillator one introduces the new complex coordinates
\be
\label{anih}
\vec a = \vec p - i \omega_1 \vec x,\qquad {\vec a}^{*}  = \vec p + i\omega_1 \vec x \quad \Rightarrow \quad H_{osc}=\frac 12 \vec{a} \cdot {\vec{a} }^{*},
\ee
and constructs $2d-1$ functionally independent integrals of motion (see, e.g., \cite{Per})
\bea
&&
I_i=a_i a^{*}_i \qquad \qquad \qquad  \text{(no sum)}
\nonumber\\[2pt]
&&
I_{1,i}=a_1 a^{*}_i+a_i a^{*}_1 \quad \quad ~ (i>1)
\eea
where $i=1,\dots,d$ is a spatial index, which all together render the model maximally superintegrable.

Consider one more isotropic oscillator with frequency $\omega_2$ and coordinates $b_i$, $b^{*}_i$ defined as in (\ref{anih}). Each oscillator entering the combined Hamiltonian
\be\label{two}
H=\frac 12 \vec{a} \cdot {\vec{a} }^{*}-\frac 12 \vec{b} \cdot {\vec{b} }^{*},
\ee
ensures $2d-1$ integrals of motion. The compound system with $2d$ configuration space degrees of freedom thus lacks for only one integral of motion to be maximally superintegrable. However, if the ratio of frequencies is a rational number $\frac{\omega_1}{\omega_2}=\frac{n_1}{n_2}$, one can construct an extra integral of motion
\be
{(a_1)}^{n_2} {(b_2)}^{n_1}+{(a^{*}_1)}^{n_2} {(b^{*}_2)}^{n_1},
\ee
which intertwines the two oscillators in (\ref{two}) and renders the full system maximally superintegrable. Obviously, this consideration can be extended to an arbitrary number of oscillators with alternating sign in the Hamiltonian. One thus concludes that (\ref{free}) is a maximally superintegrable system provided the ratios of frequencies are rational numbers $\frac{\omega _k}{\omega _{k'}}=\frac{n_k }{n_{k'}}$.
In particular, the case $\omega_k = (2k - 1)\omega_1$ with $k=1,...,n$, which is of our primary concern in this work, does belong to this class.

Note that, taking into account the form of the vector generators in (\ref{ces}) and their Fourier expansion in terms of $\cos{\left((2k-1)\omega_1 t\right)}$ and
$\sin{\left((2k-1)\omega_1t\right)}$ with $k=1,\dots,n$, one can demonstrate that the general solution (\ref{gs}) can be constructed entirely in terms of the conserved charges corresponding to the generators ${\vec C}^{(p)}$ which form a nilpotent ideal of the $l$--conformal Newton--Hooke algebra. An immediate corollary is that, within the Hamiltonian formulation for the PU oscillator, the conserved charges corresponding to the remaining generators $H$, $D$, $K$ and $M_{ij}$ as well as the superintegrals can be constructed out of ${\vec C}^{(p)}$. More details will be presented elsewhere \cite{K}. A similar role played by the vector generators in dynamical realizations of the
$l$--conformal Galilei group was revealed in \cite{GM3}.

Concluding this section, let us discuss an interplay between the $l$--conformal Newton--Hooke symmetry realized in the decoupled oscillators and symmetries which characterize each individual constituent.
As is known, dynamical symmetries of the isotropic oscillator in arbitrary dimension form the $l=\frac 12$ conformal NH group. The conserved charges (see e.g. \cite{G} and references therein)
\bea\label{l12}
&&
H=\frac 12 ({\vec p}^2+\omega^2 {\vec x}^2), \qquad
D=-\frac 12 ({\vec x} {\vec p} ) \cos{(2 \omega t)}+\frac{1}{4 \omega} ({\vec p}^2-\omega^2 {\vec x}^2) \sin{(2 \omega t)},
\nonumber\\[2pt]
&&
K=-\frac{1}{2 \omega} ({\vec x} {\vec p}) \sin{(2 \omega t)}-\frac{1}{4 \omega^2} ({\vec p}^2-\omega^2 {\vec x}^2) \cos{(2 \omega t)}+\frac{1}{2 \omega^2} H, \quad  M_{ij}=x_i p_j-x_j p_i,
\nonumber\\[2pt]
&&
{\vec C}^{(0)}={\vec p} \cos{(\omega t)}+\omega \vec{ x} \sin{(\omega t)}, \qquad  {\vec C}^{(1)}=\frac{1}{\omega} {\vec p} \sin{(\omega t)}-{\vec x} \cos{(\omega t)}
\eea
do obey the structure relations (\ref{algebra}) under the Poisson bracket, provided $R=\frac{1}{\omega}$. Note that there appears the central term in the algebra
\be
[C^{(0)}_i,C^{(1)}_j]=\delta_{ij},
\ee
which is customary for realizations of nonrelativistic conformal algebras in Hamiltonian mechanics.

For generic values of frequencies a set of $n$ decoupled isotropic oscillators in $d$ dimensions accommodates $n$ copies of the $l=\frac 12$ conformal NH algebra.
It is then interesting to see how the full $l$--conformal Newton--Hooke symmetry is accommodated in (\ref{free}) and to confront it with (\ref{l12}) realized in each constituent.
Below we do this for the simplest case of the quartic PU oscillator which corresponds to $l=\frac{3}{2}$.

In order to construct the conserved charges, one starts with the Lagrangian corresponding to the PU oscillator for $n=2$ and $\omega_2=3 \omega_1$ and builds the Noether charges associated with
the $l=\frac 32$ conformal NH symmetry. Then one uses Ostrogradski's method which yields the Hamiltonian for a generic higher derivative theory. Finally, one applies a canonical transformation which links such a Hamiltonian to that describing decoupled oscillators (for more details see, e.g., Ref. \cite{S}) and rewrites the Noether charges in terms of those variables. The result reads
\bea\label{l32}
&&
{\vec C}^{(0)}=\omega_1 \left(3 {\vec p}_1  \sin{(\omega_1 t}) - {\vec p}_2  \cos{(3 \omega_1 t)} +3 \omega_1 {\vec x}_1  \cos{(\omega_1 t)}  -
3 \omega_1 {\vec x}_2 \sin{(3 \omega_1 t)} \right),
\nonumber\\[2pt]
&&
{\vec C}^{(1)}=-{\vec p}_1  \cos{(\omega_1 t)}- {\vec p}_2  \sin{(3 \omega_1 t)}+ \omega_1 {\vec x}_1  \sin{(\omega_1 t)}
 + 3 \omega_1 {\vec x}_2  \cos{(3 \omega_1 t)} ,
 \nonumber\\[2pt]
&&
{\vec C}^{(2)}=\frac{1}{\omega_1} \left({\vec p}_1  \sin{(\omega_1 t)}+{\vec p}_2  \cos{(3 \omega_1 t)}+\omega_1 {\vec x}_1  \cos{(\omega_1 t)} + 3 \omega_1 {\vec x}_2  \sin{(3 \omega_1 t)}\right),
 \nonumber\\[2pt]
&&
{\vec C}^{(3)}=\frac{1}{\omega_1^2} \left(-3 {\vec p}_1  \cos{(\omega_1 t)} + {\vec p}_2  \sin{(3 \omega_1 t)} + 3 \omega_1 {\vec x}_1  \sin{(\omega_1 t)} - 3 \omega_1 {\vec x}_2  \cos{(3 \omega_1 t)}
 \right),
 \nonumber\\[2pt]
&&
H=\frac 12 ({\vec p}_2^2+9 \omega_1^2 {\vec x}_2^2)- \frac 12 ({\vec p}_1^2+\omega_1^2 {\vec x}_1^2), ~
K=-\frac{1}{2 \omega_1^2}  \left( A \sin{(2 \omega_1 t)}+ B \cos{(2 \omega_1 t)} - H\right),
 \nonumber\\[2pt]
&&
D=-\frac{1}{2 \omega_1} \left( A \cos{(2 \omega_1 t)}-B \sin{(2 \omega_1 t)}\right), ~ M_{ij} =x_{1 i} p_{1 j}-x_{1 j} p_{1 i}+x_{2 i} p_{2 j}-x_{2 j} p_{2 i},
\eea
where we denoted
\bea
&&
A={\vec p}_1 {\vec p}_2 - 2 \omega_1 {\vec x}_1 {\vec p}_1 - 3 \omega_1^2 {\vec x}_1 {\vec x}_2, \qquad B={\vec p}_1^2 - \omega_1^2 {\vec x}_1^2 - \omega_1 {\vec x}_1 {\vec p}_2 - 3 \omega_1 {\vec x}_2 {\vec p}_1.
\eea

It is straightforward to verify that the phase space functions (\ref{l32}) do obey the structure relations (\ref{algebra}) with $l=\frac 32$, the central terms being
\be
[C^{(0)}_i,C^{(3)}_j]=-12 \delta_{ij}, \qquad [C^{(1)}_i,C^{(2)}_j]=4 \delta_{ij}.
\ee

One thus concludes that the main difference with the $l=\frac 12$ conformal NH symmetry realized in each isotropic oscillator with the help of Eq. (\ref{l12}) is that, while the variables parametrizing the oscillators keep decoupled in the Hamiltonian and the generator of spatial rotations, they are nontrivially intertwined within all other conserved charges.
A generalization of this consideration to higher values of $l$ is straightforward albeit tedious.

\vspace{0.5cm}

\noindent
{\bf 6. Discussion}\\

\noindent

To summarize, in this work we have determined the maximal kinematical invariance group of the classical PU oscillator in arbitrary dimension.
This was demonstrated to coincide with the $l$--conformal Newton--Hooke group provided frequencies of oscillation form
the arithmetic sequence $\omega_k=(2k-1) \omega_1$, with $k=1,\dots,n$, and $l$ is the half--integer $\frac{2n-1}{2}$. The model was shown to be maximally superintegrable.
A link to $n$ decoupled isotropic oscillators was established and an interplay between the $l$--conformal Newton--Hooke symmetry and symmetries characterizing
each individual oscillator was discussed.

From the group--theoretical viewpoint, the results obtained in this paper are closely tied to the fact that the dynamical equation which describes the PU oscillator is linear. In particular, the linearity implies that the corresponding Hamiltonian is quadratic in the canonical variables. It then follows from the structure of the NH algebra (\ref{algebra}) that the generators of dilatation and the special conformal transformation are quadratic as well. The same concerns rotations. On the other hand, owing to the linear dynamics a shift of the canonical variables by an arbitrary solution to the canonical equations of motion is a symmetry which ought to be included into the full symmetry group. The generator of the shift is linear in the canonical variables and hence it does not belong to $so(2,1)\oplus so(d)$, but rather it is an element of the nilpotent subalgebra generated by $\vec C^{(p)}$.
Being expressed in terms of the canonical variables, these generators depend on time both explicitly and via the canonical variables. The latter dependence can be inferred from the Poisson bracket (cf. Eq. (\ref{algebra}))
\be
\{H,\vec{C}^{(p)}\}=A^{pq} \vec{C}^{(q)}, \qquad A^{pq}=p \delta_{p-1,q}+(p-2 l) \omega^2 \delta_{p+1,q},
\nonumber
\ee
where $p,q=0,\dots,2l$.
As a result, the dynamics of the canonical variables can be read off from the previous line. The eigenvalues of the matrix $A^{nm}$ read $\pm i\omega ,\,\pm 3i\omega ,\, ...,\,\pm 2il\omega $ \cite{GM3}. It is noteworthy that
they reproduce the constraints on the admissible values of frequencies revealed above.

Note that the conserved charges corresponding to $\vec C^{(k)}$ provide $d\times (2n)$ independent linear combinations of the canonical variables and can be viewed as initial conditions for the dynamical equations.
Hence it is no wonder that all the dynamical variables can be expressed in terms of them \cite{K}.  Worth mentioning also is that the PU model with frequencies given in Eq. (\ref{apr}) can be analyzed using the method of nonlinear realizations. It turns out, however, that this elegant method does not allow one to show in a straightforward way that the $l$--conformal NH group is the maximal symmetry group of the model \cite{K}.

Turning to possible further developments, the most urgent question is whether the $l$--conformal NH symmetry of the free PU oscillator is compatible with interactions resulting in the stable classical dynamics (cf. one--dimensional systems in \cite{S,AS,RS}). Then it would be interesting to extend the present consideration to quantum domain. In particular, it is tempting to construct a quantum counterpart of Niederer's transformation. Higher derivative dynamical systems invariant under the action of supersymmetric extensions of the $l$-conformal NH group are worth studying as well.

As we have seen above, the various methods agree on the constraint
on frequencies which guarantees the presence of the $l$--conformal NH symmetry in the PU oscillator. A clear-cut physical interpretation of those concrete values remains a challenge. Perhaps the inclusion of interaction will shed some light on the problem.

\vskip 0.5cm
\noindent
{\bf Acknowledgements}\\

\noindent
K.A and J.G are grateful to Piotr Kosi\'nski for helpful and illuminating discussions. We thank Peter Horv\'athy and Andrei Smilga for useful correspondence.
This work was supported by the NCN grant DEC-2013/09/B/ST2/02205 (K.A and J.G.) and by the RFBR grants 13-02-90602-Arm (A.G.) and 14-02-31139-Mol (I.M.).
I.M. gratefully acknowledges the support of the Dynasty Foundation.

\end{document}